# Precise Micropatterning of a Porous Poly(Ionic Liquid) *via* Maskless Photolithography for High-Performance Non-Enzymatic H$_2$O$_2$ Sensing

Ming-Jie Yin[†], Qiang Zhao*[‡,§], Jushuai Wu[†], Karoline Seefeldt[§], and Jiayin Yuan*[§,∥]

[†]Photonics Research Center, Department of Electrical Engineering, The Hong Kong Polytechnic University, Hong Kong SAR, China

[‡]Key Laboratory of Material Chemistry for Energy Conversion and Storage, Ministry of Education, School of Chemistry and Chemical Engineering, Huazhong University of Science and Technology, Wuhan 430074, China

[§]Max Planck Institute of Colloids and Interfaces, Department of Colloid Chemistry, D-14424 Potsdam, Germany

[∥]Department of Materials and Environmental Chemistry (MMK), Stockholm University, Svante Arrhenius väg 16 C, 10691 Stockholm, Sweden

**ABSTRACT:** Porous poly(ionic liquid)s (PILs) recently have been actively serving as a multifunctional, interdisciplinary materials platform in quite a few research areas, including separation, catalysis, actuator, sensor, and energy storage, just to name a few. In this context, the capability to photo-pattern PIL microstructures in a porous state on a substrate is still missing but is a crucial step for their real industrial usage. Here, we developed a method for *in situ* rapid patterning of porous PIL microstructures *via* a maskless photolithography approach coupled with a simple electrostatic complexation treatment. This breakthrough enables designs of miniaturized sensors. As exemplified in this work, upon loading Pt nanoparticles into porous PIL microstructures, the hybrid sensor showed outstanding performance, bearing both a high sensitivity and a wide detection range.

Poly(ionic liquid)s (PILs), emerging as a class of multifunctional polymers, have been receiving rapidly growing interest in fields of materials and polymer science due to a combined profile of properties and functions of two big materials families, *i.e.* ionic liquids and polymers in terms of chemical and thermal stability, non-flammability, ionic conductivity, tunable solvent-ion interaction, a wide electrochemical stability window, diverse molecular structures, surface activities and/or macromolecular processability.[1,2] PILs have been not only the target of fundamental research to expand the property window of general polymer materials, but also extensively explored for materials applications, including catalysis,[3,4] separation,[5-7] energy storage,[8-10] electrochemical devices,[11-14] sensors and actuators,[15-21] to mention a few. Among them, PIL-based sensors are a topic in fast-development because of the rich forms of interactions (ionic, covalent and non-covalent intermolecular) between the ionic liquid species in PILs and external chemical stimulus (solvents, ions, gas, radicals, *etc.*).

Lately, engineering PILs into a porous state is popularly adopted as a tandem route to expand and enhance their materials performance in various areas,[22-25] such as advanced catalysis,[3] actuators and sensors.[20,21] In this regard, PIL sensors shaped in a porous form are of particular interest because of intrinsic porous structures, *i.e.* light weight, large surface area, fast transport/diffusion, and less materials demand. To prepare porous PILs, porogenic solvent methods,[26] soft-/hard-templating approach,[27-29] and alkaline triggered complexation technique,[18,20,30,31] were introduced. The latter, *i.e.* the alkaline triggered one, is a mild and popular method because of its simplicity in operation and controllability in porous structures by the choice of building blocks[30] or applied voltage.[31]

Meanwhile, the current design of sensors, especially in industry, aims to engineering the sensing component into small or microdevices to reduce the cost. This trend of miniaturization in sensor design requires *in situ*, precise deposition of active materials on an effective area of a substrate.[32-34] For example, Javey and coworkers employed traditional photolithography technique to fabricate a wearable sensor for real-time monitoring of human sweat for health evaluation.[35] In this context, study of patterned PIL microstructures has been motivated. In 2014, Abdelhedi-Miladi *et al.* patterned azide-functionalized PILs into microstructures with the aid of photomask by UV light.[36] Later, Long *et al.* introduced a mask projection microstereolithography technique to 3D print PILs *via* photopolymerization of phosphonium polymerizable ILs.[37] More recently, we micropatterned a vinyl-functionalized PIL, poly(1-allyl-3-vinylimidazolium bromide), on the tip of an optical fiber for $CO_2$ sensing.[38] These techniques have been a significant push to advance the progress in micropatterning PILs. However, they are not universal and have limited availability of PILs of specifically designed chemical structures for these processes and add complexity.

The literature survey mentioned above logically encourage us in pursuit of a simple, fast way to micropattern PILs into a porous state to fuel further growth of the PIL sensor research. In this contribution, we report how to *in situ* photo-pattern porous PIL microstructures *via* maskless photolithography and then the consequent electrostatic complexation. This technique allows for accurate positioning of porous PILs at a microscale, which can be used in sensor design, as exemplified here with an electrode microdevice for non-enzymatic, sensitive and fast detection of $H_2O_2$.

the micropatterned PIL-MA blend film on the substrate was dried at 80 °C for 1h before immersion into a 0.2 wt % aqueous $NH_3$ solution. In this alkaline solution, MA was neutralized and deprotonated/anionized (6 carboxylate anions per molecule) by $NH_3$ to induce electrostatic complexation with the cationic PIL; meanwhile, the pores were created by a phase separation mechanism along the water diffusion into the PIL-MA blend film because of the hydrophobic nature of PIL, *i.e.* the PIL chains try to minimize their contact with water. The detailed pore formation mechanism can be referred to our previous reports.[18,20,30] Fourier transform infrared (FTIR) spectroscopy was adopted to detect the presence of deprotonated carboxylate groups, which appeared at 1560 cm$^{-1}$ in the FTIR spectrum of the $NH_3$-treated sample (Figure S1), in support of the concomitant ionic complexation in the sample.

## RESULTS AND DISCUSSION

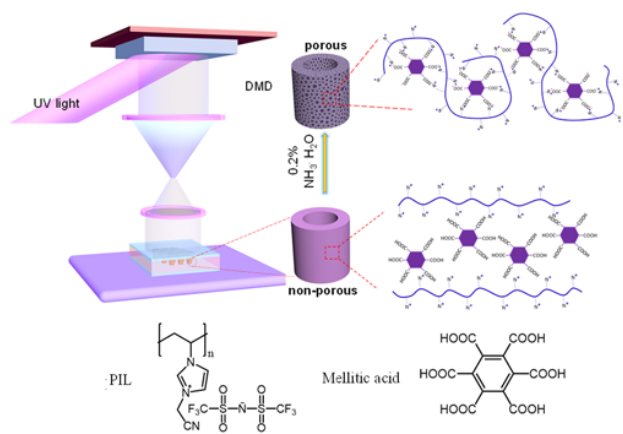

**Figure 1.** Scheme illustration of the fabrication procedure of porous PIL microstructures by maskless photolithography and the following alkaline solution treatment to introduce electrostatic complexation. The chemical structures of the PIL and mellitic acid (MA) are provided at the bottom. (DMD: digital micromirror device).

Figure 1 shows the fabrication scheme of supported porous PIL microstructures. In detail, a hydrophobic PIL with a large-sized fluorinated anion, poly[1-cyanomethyl-3-vinylimidazolium bis(trifluoromethanesulfonyl)imide], was used as the polycation,[2,3,18,39] and mellitic acid (MA) as the weak multiacid (chemical structures shown in Figure 1). They were fully dissolved in N,N-dimetylformamide (DMF) solvent before a photoinitiator (Irgacure 2959) and a crosslinker (N,N-methylenebisacrylamide) were added. The mixture solution was homogenized and then irradiated by UV light with an in-house optical maskless exposure setup,[32] which initiated polymerization of N,N-methylenebisacrylamide that crosslinked only the illuminated area, a key step in forming the micropattern (It should be emphasized that the PIL/MA blend pattern is stable in the development solvent only with adding enough crosslinkers (>1.5 wt %)). After development using a DMF/isopropanol (v/v = 1:3) mixture solution that could quickly wash away polymers in the non-crosslinked area,

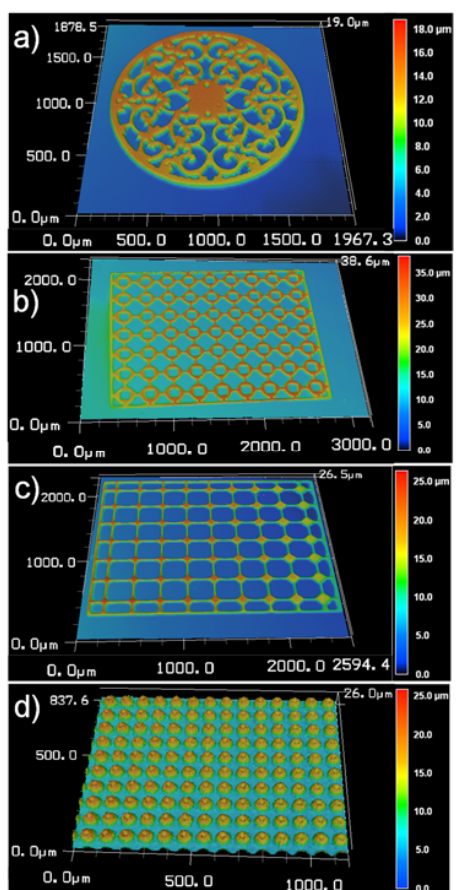

**Figure 2.** Laser scanning confocal images of the fabricated porous PIL micropatterns: (a) a pattern of a traditional Chinese paper-cut; (b) micro-coin array; (c) gradient-dotted line; (d) micro-tube array.

Figure 2 illustrates the laser scanning confocal images of four cases of porous PIL micropatterns produced in this work. Figure 2a is a pattern of a traditional Chinese paper-cut. The sophisticated image proves the success and robustness of our maskless photolithography technique towards micropatterning of PILs. In parallel, a regular micro-coin array, more complex gradient-dotted lines, and a micro-tube array are shown in Figure 2b-d, respectively, which demonstrate the exquisite, flexible and quantitative control of micro-elements of different morphology, height and size in these patterns. Worth to mention is that all these patterns are completed within merely 30 s of a UV irradiation process at a routine power density of 99.13 mW/cm$^2$, indicating a convenient, swift and general technique. The detailed geometric parameters of these PIL microstructures are provided in Figure S2. It should be noted that the MA-PIL blend microstructures were dried before immersing into a 0.2 wt % aqueous NH$_3$ solution, thus one might expect if this step should affect the resolution of the final porous microstructure. By measuring the line width change before and after drying the micropatterns, the shrinkage is determined to be 5.8 ± 1.2% (Figure S3). This weak shrinkage is understandable because the majority of solvent was readily evaporated during UV irradiation process and there is little residual solvent in the final micropatterned structure, in accordance with the observation that the drying process actually has little effect on the final microstructures. To be stressed is that the smallest line width that can be reached in such experiments is 5.16 μm, as shown in Figure S4.

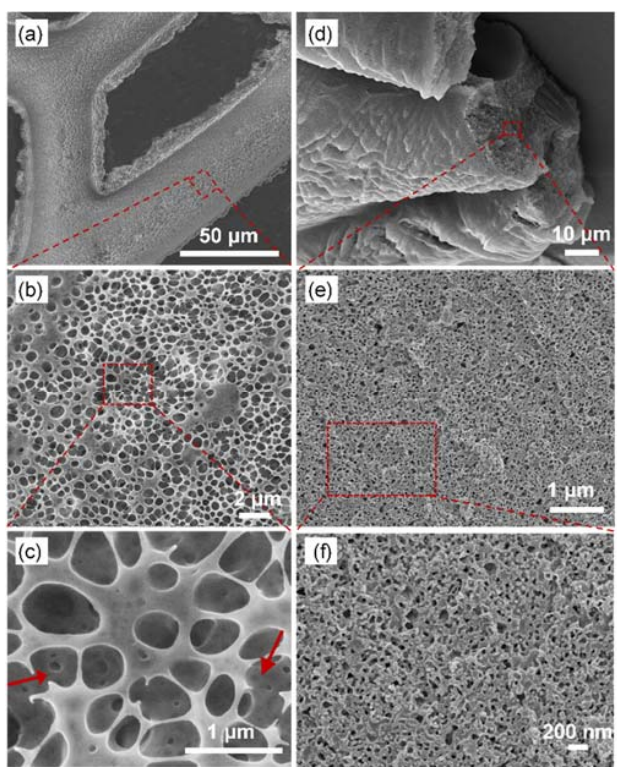

**Figure 3.** SEM images of micropatterned porous PILs: (a)-(c) are surface morphologies with different magnifications; (d)- (f) are their cross-section views, respectively.

Our goal in this work is not merely to produce the maskless micropatterns of PILs but also porous structures that are challenging and add extra complexity. Representative scanning electron microscopy (SEM) images of these micropatterned PILs were displayed in Figure 3. Figure 3a is an overview of the flat surface of the microstructure, while in Figure 3b uniform and densely packed pores of 300 ± 50 nm in average size are clearly visible. The cross-sectional SEM images were taken (Figure 3d-3f), whereby abundance of nanopores (10~100 nm) are unambiguously observed (Figure 3f), indicative of hierarchical porous nature of the PIL micropatterns. This nano-/macro pore system facilitates high-performance microdevice development, *e.g.* sensors and batteries.[1] Besides, the influence of crosslinker concentration on the porous structure was also studied (Figure S5). By increasing the crosslinker concentration from 1.5 wt % to 4.4 wt %, the pore density decreased rapidly and the pattern became dense in the end. This is likely because the excessive crosslinking inhibited the PIL chain mobility and phase separation, thus depressing the pore formation.

The charged, porous framework of PILs offers a large surface to accommodate metal nanoparticles to introduce sensing function to task-specific microdevices.[3,24] In this work, platinum nanoparticles (PtNPs) were loaded by immersing the micropattern into the pre-synthesized PtNPs dispersion, during which the negatively charged PtNPs are electrostatically immobilized onto the PIL micropattern surface. PtNPs are known to catalyze many reactions, a famous one of which is the decomposition of $H_2O_2$ into $H_2O$ and oxygen, as shown in Figure 4a. In this reaction, an electron transfer process occurs simultaneously, resulting in the rise of current density when it is electrochemically monitored. $H_2O_2$, a product existing in biochemical processes in living organisms, plays a key role in normal cellular function and proliferation.[40-44] For example, it regulates the DNA damage and produces intracellular thermogenesis, both of which are the basic process for normal life.[45-47] Detection of $H_2O_2$ in a quantitative manner of high sensitivity and low detection limit in a reasonably broad range is of high importance, and it is our interest to employ it here as a model reaction to demonstrate the utilization of the sensor device built up from our maskless photo-deposited porous PILs. To fabricate the PIL-Pt hybrid sensor, the porous PILs were first micropatterned on an indium tin oxide (ITO) electrode with a flexible substrate of polyethylene terephthalate (PET). The transmission electron microscopy (TEM) image in Figure 4b shows uniformly sized PtNPs, the average size of which is 2.3 ± 0.4 nm, as determined statistically in the inset of Figure 4b. The high-resolution TEM (HR-TEM) lattice image of the PtNPs in Figure 4c visualizes the lattice spacing of 0.24 nm, a value in consistency with that of Pt (111) planes.[48,49]

The microfabricated $H_2O_2$ sensor was tested by immersing the sensor in a phosphate-buffered saline (PBS) solution at different concentration of $H_2O_2$ ($C_{H2O2}$). Figure 4d is a typical time-dependent current response curve of the electrode at -0.56 V along successive addition of $H_2O_2$. The current increases immediately upon the addition of $H_2O_2$ solution, and rapidly reaches a steady state. The response time is as short as 18 s (to reach 90% of the steady state current), which originates from the porous structure of PILs which facilitates the high-speed transportation of molecules and electrons.[18,50] In a control experiment, a $H_2O_2$ sensor made from non-porous micropatterned PILs was also tested, the response time of which is more than 30s (Figure S7b).

**Table 1** Comparison of analytical performance of $H_2O_2$ sensors based on Pt catalysis.

| Electrode materials | Sensitivity ($\mu A/cm^2/mM$) | Linear range (mM) | Detection limit ($\mu M$) | Ref. |
| --- | --- | --- | --- | --- |
| PtNPs@UiO-66/GCE | 75.33 | 0.005-14.75 | 3.06 | 51 |
| Pt/rGO–CNT | 1.41 | 0.0001-0.025 | 0.01 | 52 |
| Pt–MnO$_2$/rGOP | 129.5 | 0.002-13.33 | 1 | 53 |
| rGO – Pt | 459 | 0.00005-3.475 | 0.2 | 54 |
| PVA–MWCNTs–PtNPs/GCE | 122.63 | 0.002-3.8 | 0.7 | 55 |
| BN–Pt/GCE | 3.4 | 0.1-1.2 | 0.11 | 56 |
| PtNPs-CNF-PDDA/SPCEs | 46.7 | 0.0025-10 | 2.4 | 57 |
| SPGFE/MWCNTC/PtNP | - | 0.005-2 | 1.23 | 58 |
| PtNPs/PILs complex/ indium tin oxide (ITO) (porous) | 301.75 | 0.005-45 | 0.2 | This work |
| PtNPs/PILs complex/ITO (nonporous) | 57.38 | 0.005-50 | 5.8 | This work |

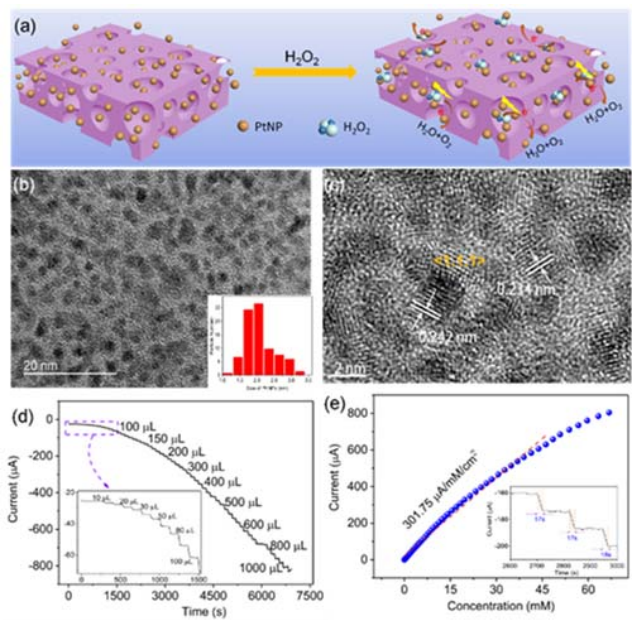

**Figure 4.** (a) Scheme illustration of the decomposition of $H_2O_2$ by the PIL-Pt hybrid sensor. (b) TEM image of the PtNPs. The inset is a statistical size distribution curve. (c) HR-TEM image of the PtNPs. (d) Amperometric response of the PIL-Pt hybrid electrode along successive addition of $H_2O_2$ (0.4 M) with different volume in 80 mL PBS buffer solution (pH = 7.4) at an applied potential of 0.56 V. The inset is a magnified image in the 0 - 1500s range. (e) The steady state current response to the change of $H_2O_2$ concentrations.

Figure 4e plots the current as a function of $C_{H2O2}$. The steady-state current of the $H_2O_2$ sensor increases linearly with $C_{H2O2}$ up to 45 mM. The calculated sensitivity is as high as 301.75 $\mu A \cdot mM^{-1} \cdot cm^{-2}$, ranked in the top class of PtNPs-based $H_2O_2$ sensors (Table 1). Its sensitivity is 6 times higher than that of the referenced nonporous sensor, which is logically attributed to the porous nature of the patterned PILs that can accommodate more PtNPs.[49] Besides, the $H_2O_2$ sensing microdevice offers a low detection limit of 0.2 μM at a signal-to-noise ratio of 3 (the noise level of the baseline is 2.01×10$^{-5}$), which stands for the top-quality of PtNPs-based $H_2O_2$ sensors.[53-55,57,58] It should be mentioned that though the detection limit is higher than the Pt/rGO–CNT (0.01 μM)[52] and PVA–MWCNTs–PtNPs/GCE (0.11 μM)[56], the PIL-Pt hybrid sensor benefits from a significantly larger linear detection range up to 45 mM.

## CONCLUSIONS

In summary, a breakthrough in micropatterning PILs with dense pores *via* a fast, maskless photolithography approach combined with a following alkaline treatment was reported here. Being simultaneously microstructured and porous, the PIL micropatterns after loading with PtNPs were readily engineered onto an electrode and served as a hybrid $H_2O_2$ sensor microdevice. A high linear sensitivity towards $H_2O_2$ up to 45 mM, and a fast response were achieved simultaneously. Giving the diverse function scope of PILs, the photoprinting property of the porous PILs is useful and expected to enable many other microdevice applications, such as soft robotics and wearable healthcare devices.

## MATERIALS AND METHODS

**Materials.** 2-Hydroxy-4'-(2-hydroxyethoxy)-2-methylpropiophenone (Irgacure 2959), N,N-methylenebisacrylamide (MBA), isopropyl alcohol (IPA), and lithium bis(trifluoromethanesulfonyl)imide (LiTf2N, 99.95%) were purchased from Sigma-Aldrich. Mellitic acid (MA) and N,N-dimethylformamide (DMF) were obtained from J&K Scientific Ltd. (China). $H_2PtCl_6$ (99%), $NH_3·H_2O$ (25%), polyvinylpyrrolidone (PVP), and $NaBH_4$ were purchased from Shenzhen Chemical Reagent Company (China). Poly(3-cyanomethyl-1-vinylimidazolium bromide) (termed PCMVImBr) was synthesized according to a previous method.[59] Poly(3-cyanomethyl-1-vinylimidazolium bis(trifluoromethanesulfonyl)imide) was obtained via anion exchange of PCMVImBr with LiTf$_2$N. Deionized (DI) water with a resistance of 18 MΩ cm was used in all experiments.

**Micropatterning of porous PIL microstructures.** 0.3 g of PIL and 0.06 g of MA were dissolved in 6 g of DMF under stirring. Then, 0.06 g of Irgacure 2959 and 0.1 g of MBA were added stepwise until fully dissolved under stirring. The prepared photoresist solution was stored for use.

A home-made maskless micropatterning set-up was adopted for microfabrication process. A flexible substrate of polyethylene terephthalate (PET) deposited with indium tin oxide (ITO) electrodes was used for microstructure fabrication. The substrate was placed on a glass slide. The pre-designed microstructures were converted into own-defined image data and then loaded onto the digital-mirror device chip for optical patterns generation. UV light source (365 nm) with an intensity of 99.13 mW/cm$^2$ was employed for micropatterning. The exposure time was between 20 and 30 s, depending on the microstructures fabricated. The microstructures were developed by a DMF/IPA (v/v=1:3) mixed solvent, which took 30 s to completely wash the unexposed part. Next, the fabricated microstructures were heated at 80 °C for 1 h and immersed into 0.2 wt% aqueous $NH_3$ for 1 h, washed with DI water and directly used for PtNPs loading without further drying.

**PtNPs synthesis.** The synthesis of PtNPs follows a typical $NaBH_4$ reduction process: $H_2PtCl_6$ was dissolved in 200 mL DI water with a concentration of 1 mM. Next, 0.35 g of PVP was added into the solution under stirring until fully dissolved. Then, $NaBH_4$ solution (50 mM, 50 mL) was slowly dropped into the above solution under vigorous agitation, with the color gradually changed to dark brown and bubbles generating. The solution was stirred for 2 h, and PtNPs with a concentration of 0.8 mM was obtained.

**Characterizations.** Microstructures of PILs were measured by 3D laser scanning confocal microscope (VK-X200, KEYENCE, Japan) and field emission scanning electron microscopy (JEOL Model JSM-6490). The former is a non-contact laser scanning imaging machine. The magnification of lens used for scanning was 50×. Transmission electron microscope (JEOL Model JEM-2011) was used for characterizing the morphology and sizes of PtNPs

**Microfabrication and tests of $H_2O_2$ sensors.** The prepared photoresist solution was dropped onto an ITO/PET substrate and micropatterned with an area of 5 mm$^2$ via the above-mentioned procedure. One sensor was treated with a 0.2 wt% $NH_3$ aqueous solution for 1 h (porous); another was treated only with DI water for 1 h (non-porous). Finally, the sensors were immersed into PtNPs dispersion for 2 h.

The electrochemical tests were performed on a PalmSens3 electrochemical workstation (PalmSens, The Netherlands). A three-electrode cell with a sample volume of 80 mL was employed. The platinum electrode and an Ag/AgCl electrode were employed as the counter electrode and the reference electrode, respectively. Amperometric detection was performed under an applied potential of -0.56 V.

## ASSOCIATED CONTENT

**Supporting Information**.
Additional experimental results, including FTIR spectrum, geometric parameters of the fabricated porous PIL microstructures, TEM image of PtNPs in the PIL-MA porous film and other control experiments. This material is available free of charge via the Internet at http://pubs.acs.org.

## AUTHOR INFORMATION

**Corresponding Author**
* Q.Z.: zhaoq@hust.edu.cn; J.Y.: jiayin.yuan@mmk.su.se
**Author Contributions**
All authors have given approval to the final version of the manuscript.

## ACKNOWLEDGMENT

This work was partially supported by Germany/Hong Kong Joint Research Scheme (Grant No.: G-PolyU505/13), German Academic Exchange Service (DAAD, Grant No.: 57054931), PolyU Strategic Development Special Project (Grant No.: 1-ZVGB), and the ERC Starting Grant 639720 NAPOLI and the Wallenberg Academy Fellow program (KAW 2017.0166) from the Knut and Alice Wallenbergs Foundation. The authors greatly thanked Prof. A.P. Zhang at Hong Kong Polytech University for his support in lab instruments and research discussions.

# Supplementary Information

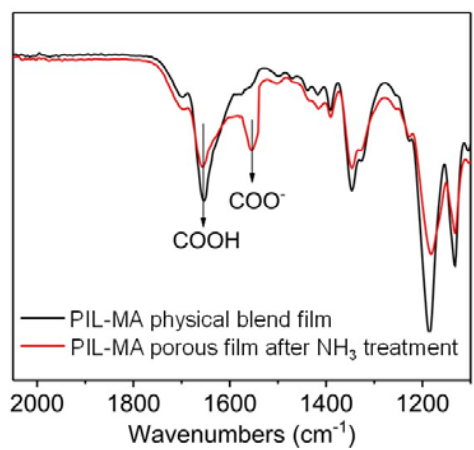

**Figure S1.** FT-IR spectra of the physical blend film of PIL and MA (black) and the PIL-MA porous film after immersion in a 0.2 wt % aqueous $NH_3$ solution.

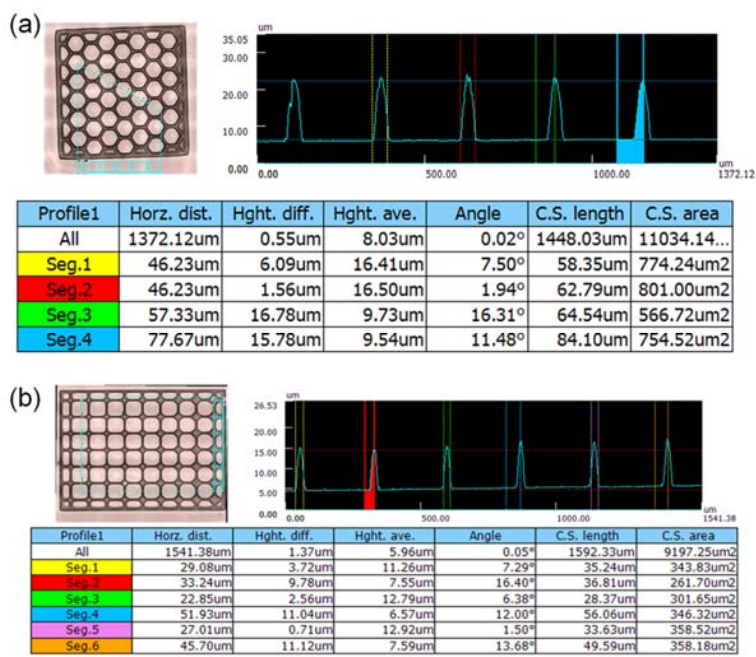

**Figure S2.** Measured geometric parameters of the fabricated porous PIL microstructures: (a) honeycomb; (b) gradient-dotted line.

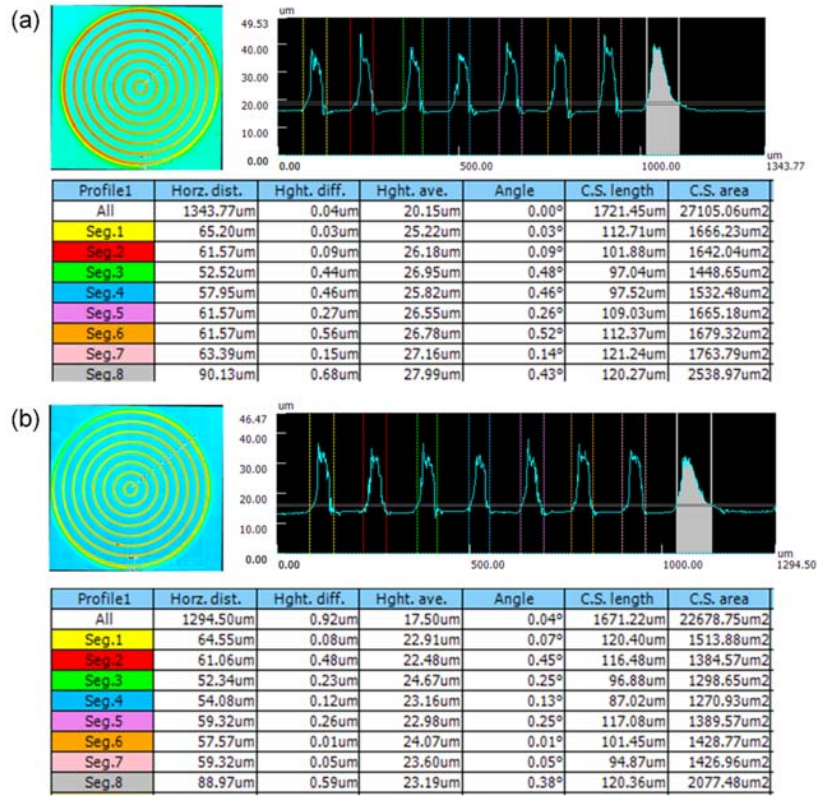

**Figure S3.** Measured geometric parameters of the fabricated PIL microstructures before (a) and after (b) the drying process.

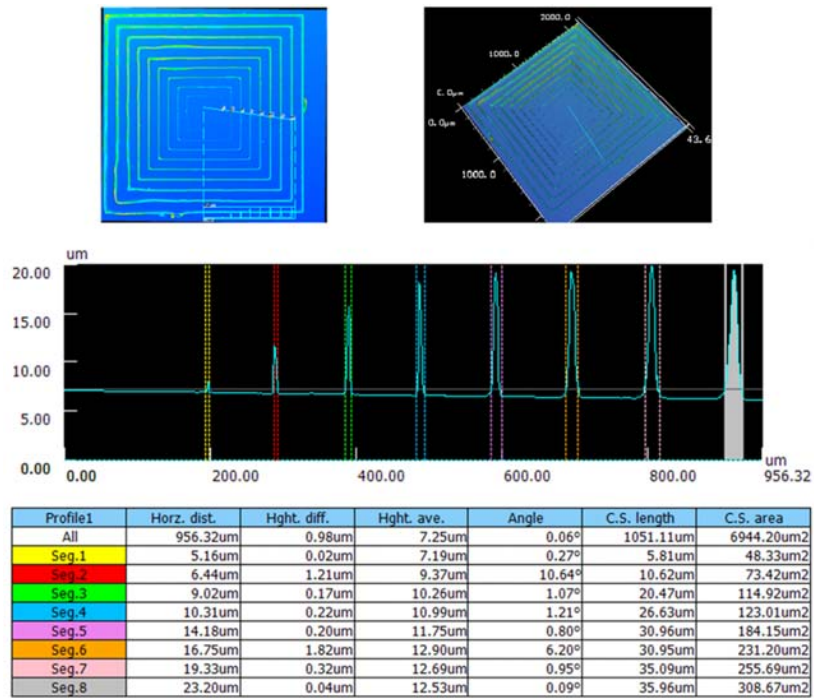

**Figure S4.** Images and line-width parameters of a concentric-square PIL microstructure measured by a 3D laser scanning confocal microscope.

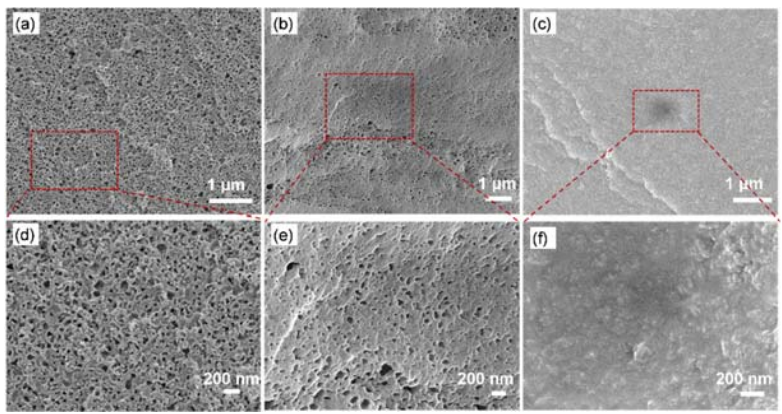

**Figure S5.** Cross-section SEM images of PIL complex with different crosslinker concentrations: (a) 1.5 wt%; (b) 3 wt%; and (c) 4.4 wt%, respectively. (d)-(f) are their corresponding enlarged images.

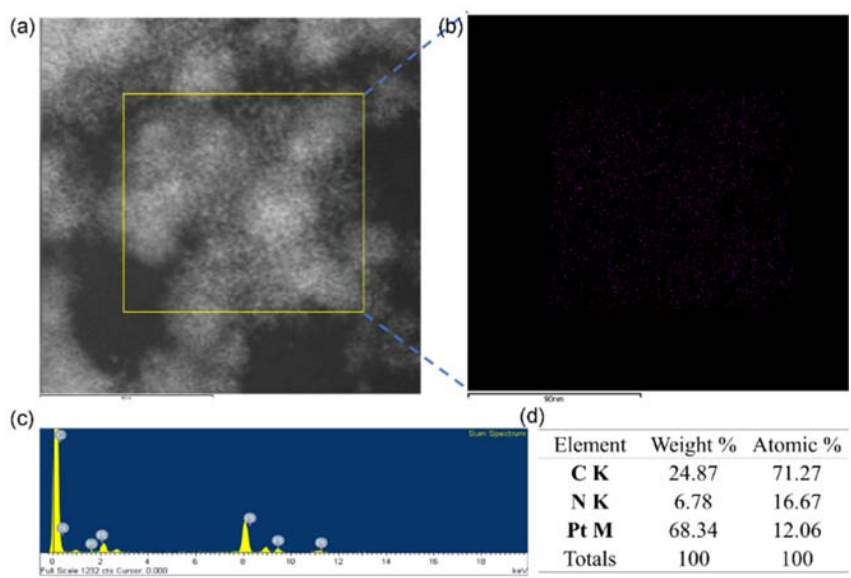

| Element | Weight % | Atomic % |
|---|---|---|
| C K | 24.87 | 71.27 |
| N K | 6.78 | 16.67 |
| Pt M | 68.34 | 12.06 |
| Totals | 100 | 100 |

**Figure S6.** (a) TEM image of PtNPs in the PIL-MA porous film; (b) Elemental mapping of Pt and (c) the corresponding EDS spectrum; (d) main elements weight ratio in the sample.

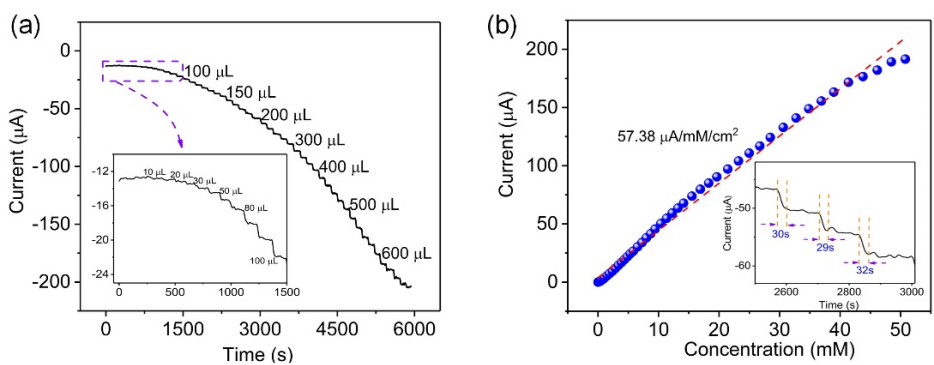

**Figure S7.** Performance of microfabricated $H_2O_2$ sensor with non-porous PIL-Pt sensor electrode tested in PBS (pH = 7.4) solution (a, b).